\RequirePackage{fix-cm}
\documentclass{svjour3}                     % onecolumn (standard format)
\smartqed  % flush right qed marks, e.g. at end of proof
\usepackage{graphicx}
\usepackage{bm}
\usepackage{amsmath}
\usepackage{amssymb}
\usepackage{mathptmx}      % use Times fonts if available on your TeX system
\usepackage{latexsym}
\usepackage{verbatim}
\usepackage{color}
\usepackage[english]{babel}
\journalname{Journal of Computational Electronics}
\begin{document}
%single-molecule nanojunctions
\title{First-principles quantum transport modeling of thermoelectricity in single-molecule nanojunctions with graphene nanoribbon electrodes}

%\subtitle{Do you have a subtitle?\\ If so, write it here}

\titlerunning{First-principles modeling of thermoelectricity in single-molecule nanojunctions}        % if too long for running head

\author{Branislav K. Nikoli\' c \and
        Kamal K. Saha \and
        Troels Markussen \and
        Kristian S. Thygesen
}

%\authorrunning{Short form of author list} % if too long for running head

\institute{Branislav K. Nikoli\' c  \and
           Kamal K. Saha  \at
           Department of Physics and Astronomy, University of Delaware, Newark, DE 19716, USA  \\
              \email{bnikolic@udel.edu}           %  \\
%             \emph{Present address:} of F. Author  %  if needed
           \and
           Troels Markussen \and Kristian S. Thygesen  \at
           Center for Atomic-scale Materials Design (CAMD), Department of Physics, Technical University of Denmark, DK-2800 Kongens Lyngby, Denmark
}

\date{Received: date / Accepted: date}
% The correct dates will be entered by the editor

\maketitle

\begin{abstract}
We overview nonequilibrium Green function combined with density functional theory (NEGF-DFT) modeling of independent electron and phonon transport
in nanojunctions with applications focused on a new class of thermoelectric devices where a single molecule is attached to two metallic zigzag graphene nanoribbons (ZGNRs) via highly transparent contacts. Such contacts make possible  injection of evanescent wavefunctions from ZGNRs, so that their overlap within the molecular region generates a peak in the electronic transmission.  Additionally, the spatial symmetry properties of the transverse propagating states in the ZGNR electrodes {\em suppress} hole-like contributions to the thermopower. Thus optimized thermopower, together with diminished phonon conductance through a ZGNR$|$molecule$|$ZGNR inhomogeneous structure, yields the thermoelectric figure of merit $ZT \sim 0.5$ at room temperature and $0.5 < ZT < 2.5$ below liquid nitrogen temperature. The reliance on evanescent mode transport and symmetry of propagating states in the electrodes makes the electronic-transport-determined power factor in this class of devices largely insensitive to the type of sufficiently short conjugated organic molecule, which we demonstrate by showing that both 18-annulene and C10 molecule sandwiched by the two ZGNR electrodes yield similar thermopower. Thus, one can search for molecules that will further reduce the phonon thermal conductance (in the denominator of $ZT$) while keeping the electronic power factor (in the nominator of $ZT$) optimized. We also show how often employed Brenner empirical interatomic potential for hydrocarbon systems fails to describe phonon transport in our single-molecule nanojunctions when contrasted with first-principles results obtained via NEGF-DFT methodology.

% \subclass{MSC code1 \and MSC code2 \and more}
\end{abstract}

\section{Introduction}\label{sec:intro}

\subsection{Why study nanoscale thermoelectrics?}\label{sec:why}

Thermoelectrics transform temperature gradients into electric voltage and vice versa. Although a plethora of widespread applications has been envisioned, their usage is presently limited by their small efficiency~\cite{Vining2009}. Thus, careful tradeoffs are required to optimize the dimensionless figure of merit
\begin{equation}\label{eq:zt}
ZT=S^2GT/\kappa,
\end{equation}
which quantifies the maximum efficiency of a thermoelectric cycle conversion in the linear-response regime. This is due to the fact that $ZT$ contains unfavorable combination of the thermopower $S$, average temperature $T$, electrical conductance $G$ and thermal conductance $\kappa =\kappa_{\rm el} + \kappa_{\rm ph}$. The total thermal conductance has contributions from both electrons $\kappa_{\rm el}$ and phonons $\kappa_{\rm ph}$. The devices with $ZT > 1$ are regarded as good thermoelectrics, but values of $ZT > 3$  are required for thermoelectric devices to compete in efficiency with mechanical power generation and refrigeration~\cite{Vining2009}.

The traditional efforts to increase $ZT$ have been directed toward selective reduction of the lattice thermal conductivity $\kappa_{\rm ph}$,  using either complex (through disorder in the unit cell) bulk materials~\cite{Snyder2008} or bulk nanostructured materials~\cite{Minnich2009}, while at the same time maintaining as optimal as possible electronic properties encoded in the power factor $S^2G$. However, decennia of intense research along these lines have increased $ZT$ of bulk materials only marginally~\cite{Vining2009}. A complementary approach engineers  electronic density of states to obtain a sharp singularity~\cite{Minnich2009,Mahan1996} near the Fermi energy $E_F$ which can enhance the power factor $S^2G$, such as in Tl-doped PbTe where $ZT \sim 1.5$ has been reached at 775 K~\cite{Heremans2008}.

The nanoscale and low-dimensional~\cite{Kim2009} devices offer additional degrees of freedom that can be tailored to achieve high-$ZT$, as exemplified by the recent ground-breaking experiments demonstrating how rough silicon nanowires (SiNW) can act as efficient thermoelectrics ($ZT \simeq 0.6$ at $T=300$ K) although bulk silicon ($ZT=0.01$ at $T=300$ K) is not~\cite{Hochbaum2008,Boukai2008}. Another example of nanoscale thermoelectrics has emerged from the recent experiments measuring thermopower of single-molecule nanojunctions~\cite{Reddy2007,Baheti2008,Malen2009,Malen2010,Tan2010} and quantum dots~\cite{Hoffmann2009}.

Availability of efficient nanoscale thermoelectrics could make possible targeted cooling of local hotspots~\cite{Chowdhury2009} due to the ease of on-chip integration. To make use of low-dimensional thermoelectric devices in macroscale applications, many nanowires must be placed in parallel, so issues of the nanowire size and packing density arise~\cite{Kim2009}.

Besides device applications, the search for optimal $ZT$ has ignited basic research in condensed matter physics and various engineering disciplines aimed at deepening our understanding of heat flow in {\em nanoscale}  or in {\em unconventional} bulk systems. For example, the recent review article~\cite{Dubi2011} on heat flow and thermoelectricity in single-molecule nanojunctions and atomic wires highlights that even apparently basic issues are not well understood in such systems. In conventional systems, where heat and charge currents are transported by Landau quasparticles, $ZT$ is normally limited by the Wiedemann-Franz law stating that $\kappa_{\rm el}/GT$ is a system-independent constant.\footnote{We should mention here that the Lorenz ratio  $\kappa_{\rm el}/GT$ calculated for transport of {\em non-interacting} electrons through several single-molecule nanojunctions shows variations by tens of percent from the Wiedemann-Franz law as the chemical potential crosses a transmission resonance, and much larger deviation around the transmission nodes~\cite{Bergfield2009}.} However, the Wiedemann-Franz law is a result of Fermi-liquid theory and breaks down~\cite{Kubala2008} in correlated bulk materials or in nanoscale systems (such as quantum dots or metallic islands) with strong Coulomb interaction effects. This has necessitated the development of novel theoretical techniques~\cite{Held2009} to tackle thermoelectricity in  correlated bulk materials, such as the combination of the dynamical mean field theory~\cite{Held2007} with the local density approximation and the Kubo formula, which has revealed enhanced thermopower in, e.g., FeSb$_2$ and Na$_x$CoO$_2$ due to electronic correlations~\cite{Wissgott2010}. In the realm of nanoscale correlated systems, significant $ZT$ values were predicted for, e.g.,  Kondo correlated quantum dots~\cite{Boese2001}, metallic single-electron transistors~\cite{Kubala2008} and Kondo insulator nanowires~\cite{Zhang2011}.

\subsection{What is interesting about thermoelectricity in single-molecule nanojunctions?}

Very recent experiments~\cite{Reddy2007,Baheti2008,Malen2009,Malen2010,Tan2010} have measured thermopower $S$ as induced thermoelectric voltage in response to a temperature difference across organic molecule sandwiched between two gold electrodes. This has ignited vigorous theoretical efforts~\cite{Dubi2011,Bergfield2009,Pauly2008,Ke2009,Finch2009,Liu2009,Liu2009a,Liu2011a,Quek2011,Nozaki2010,Saha2011,Sergueev2011,Bergfield2011,Murphy2008,Leijnse2010,Entin-Wohlman2010,Stadler2011} to explore devices where a single organic molecule is attached to metallic or semiconducting~\cite{Nozaki2010} electrodes. In such nanojunctions, the dimensionality reduction and possibly strong electronic correlations~\cite{Bergfield2011,Murphy2008,Leijnse2010} allow for the increase in $S$ concurrently with diminishing $\kappa_{\rm ph}$ while keeping the nanodevice disorder-free~\cite{Markussen2009}. For example, creation of sharp transmission resonances near the Fermi energy $E_F$ by tuning the {\em chemical properties} of the molecule and molecule-electrode contact via chemical functionalization can substantially enhance~\cite{Finch2009,Nozaki2010} the thermopower $S$ which depends on the derivative of the conductance near $E_F$. At the same time, the mismatch in the phonon density of states at the electrode$|$molecule interface can severely disrupt phonon transport~\cite{Tsutsui2010a}, when compared to homogenous clean electrode made of the same material, thereby leading to small $\kappa_{\rm ph}$.

Besides offering a new route toward high-$ZT$ devices, thermoelectric properties of single-molecule nanojunctions have been investigated as a tool that could resolve a number of fundamental issues in molecular electronics~\cite{Song2011}. For example, Ref.~\cite{Paulsson2003} has suggested that thermoelectric voltage in single-molecule nanojunctions, which is large enough and rather insensitive to the details of coupling to the contacts, can be employed to locate the position of the Fermi energy relative to the molecular levels. This concept has sparked the development of  experimental techniques that can unambiguously identify the molecular orbital closest to the Fermi level of the electrodes by measuring thermopower and current-voltage characteristics of single-molecule nanojunctions~\cite{Tan2010}. The measurement of $S$ could also probe room-temperature quantum interference effects in transport through single-molecule nanojunctions~\cite{Bergfield2011,Cardamone2006,Ke2008,Saha2010,Markussen2010a,Markussen2011}  where Ref.~\cite{Bergfield2009} has predicted dramatic enhancement of the thermopower near a transmission node because the flow of entropy (as an inherently incoherent quantity) is not blocked by destructive quantum interference which block the charge current and lead to vanishing conductance.

The single-molecule nanojunctions also provide access to nonlinear thermoelectric properties. Most theoretical studies of nanoscale thermoelectrics have focused on the linear-response  regime (i.e., close to equilibrium) where one operates close to the small voltage $V = - S \Delta T$ which exactly cancels the current induced by the small temperature bias $\Delta T$. As $ZT\rightarrow \infty$, the efficiency approaches the ideal Carnot value $\eta \rightarrow \eta_C = 1 - T/(T+\Delta T)$. However, in the linear-response regime $\Delta T \ll T$ typically investigated for bulk materials, the efficiency stays low $\eta \approx \Delta T/T$ even if $ZT$ can be made very large. The recent experiments~\cite{Reddy2007} on single-molecule  nanojunctions showed nonlinearities in the measured $S$ already at $\Delta T \approx 0.1 T$. Furthermore, the analysis of simple phenomenological models of nanojunctions has suggested~\cite{Galperin2008,Leijnse2010} that optimal thermoelectric operation can be achieved in the out-of-equilibrium nonlinear regime. Such regime also requires theoretical approaches that go beyond usual concepts of the figure of merit $ZT$ and thermopower $S$ defined in the linear-response regime~\cite{Leijnse2010}.

\begin{figure}
\begin{center}
\includegraphics[scale=0.5,angle=0]{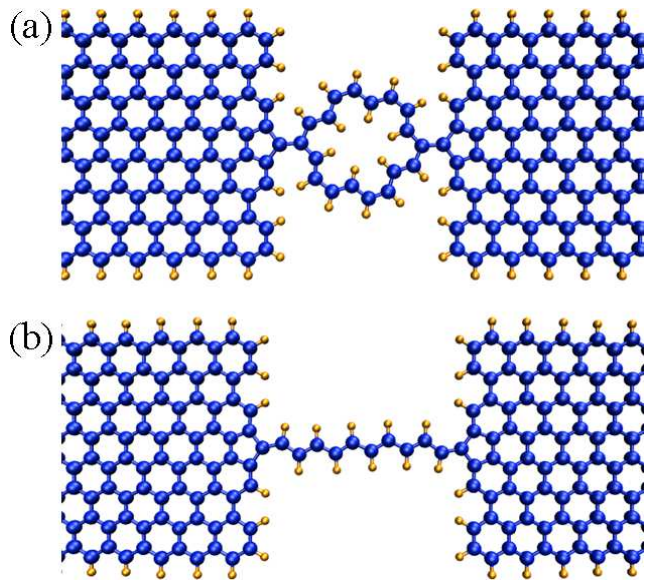}
\end{center}
\caption{Schematic view of (a) ZGNR$|$18-annulene$|$ZGNR and (b) ZGNR$|$C10$|$ZGNR single-molecule nanojunctions.
The contact between the source and the drain 8-ZGNR (consisting of eight zigzag
chains) metallic electrodes and molecules is made via five-membered rings of carbon atoms
(dark blue). The hydrogen atoms (light yellow) are included to passivate the edge carbon atoms in the nanoribbon or dangling bonds in the molecules.}
\label{fig:fig1}
\end{figure}

\subsection{What is the scope of this mini-review?}\label{sec:overview}

This article has two principal goals: ({\em i}) to present some technical details on the usage of nonequilibrium Green function combined with density functional theory (NEGF-DFT) computational methodology to study electron and phonon transport quantities which are required to understand thermoelectricity in  single-molecule nanojunctions; ({\em ii}) to overview applications of NEGF-DFT methodology to a new class of single-molecule-based thermoelectrics~\cite{Saha2011} illustrated in Fig.~\ref{fig:fig1}.

In these devices, organic molecule is attached to metallic graphene nanoribbons with zigzag edges (ZGNR) via strong covalent bond which provides high transparency of ZGNR$|$molecule contact. This allows for evanescent wavefunctions to penetrate from the electrodes into the molecular region thereby generating a transmission resonance which, together with spatial symmetry of the transverse propagating states within the ZGNR electrodes, yields  highly optimized power factor $S^2G$. Using two different molecules in Fig.~\ref{fig:fig1}, we demonstrate that $S^2G$ optimized in this fashion is independent on the type of the molecule employed, which is a feature that can be exploited to search for the molecule that will bring the largest reduction of the phonon thermal conductance.

The article is organized as follows. In Sec.~\ref{sec:negfe} we discuss NEGF-DFT approach to the computation of electronic conductance, thermopower and thermal conductance for realistic single-molecule nanojunctions built from carbon and hydrogen atoms. The same approach applied to phonon thermal conductance is discussed in Sec.~\ref{sec:negfph} for the elastic regime where electron-phonon~\cite{Frederiksen2007,Dash2011}  interactions or phonon-phonon~\cite{Mingo2006} scattering processes are neglected. The electronic transmission function and thermopower for ZGNR$|$18-annulene$|$ZGNR and ZGNR$|$C10$|$ZGNR junctions are discussed and compared in Sec.~\ref{sec:electron}. In Sec.~\ref{sec:symmetry} we show how the usage of ZGNR electrodes, whose transverse propagating eigenstates have special symmetry properties in real space, makes it possible to block the hole-like portion of the transmission function thereby enhancing the thermopower.  Section~\ref{sec:phonon} shows the phonon transmission function and the corresponding thermal conductance for these two junctions. In Sec.~\ref{sec:phonon}, we also compare phonon transport quantities obtained via full NEGF-DFT methodology to computationally faster NEGF coupled to Brenner empirical interatomic potential (EIP) where we find failure of the latter technique to describe nanojunctions in Fig.~\ref{fig:fig1}. Using the quantities computed in Sec.~\ref{sec:electron} and Sec.~\ref{sec:phonon}, we construct the thermoelectric figure of merit $ZT$ in Eq.~\eqref{eq:zt} for the two single-molecule nanojunctions in Sec.~\ref{sec:zt}. We summarize and discuss briefly future directions for the computational modeling of single-molecule thermoelectrics in Sec.~\ref{sec:conclusion}.

\section{NEGF-DFT methodology for the computation of electron transport-determined thermoelectric quantities}\label{sec:negfe}

The traditional tools for the computation of thermoelectric transport coefficients, such as the semiclassical Boltzmann equation~\cite{Vo2008}, {\em cannot} be used for quasiballistic nanometer-size active region attached to much larger reservoirs.\footnote{For comparison between Boltzmann semiclassical and Landauer quantum transport approaches applied to thermoelectric transport coefficients of conventional translationally invariant systems see Ref.~\cite{Jeong2010}.} The proper description of such {\em open quantum system} can be achieved using quantum master equations for the reduced density matrix of the active region~\cite{Breuer2002,Mitra2004,Dubi2008,Timm2008} or NEGF formalism~\cite{Haug2007}. The former is typically used when the active region is weakly coupled to the reservoirs (so that coupling between the molecule and the electrodes is treated perturbatively), while the latter is employed in the opposite limit~\cite{Haupt2010}. The complementary nature of the two methods and the boundaries of their validity  were analyzed in, e.g., Ref.~\cite{Hartle2009} using simple phenomenological models of single-molecule nanojunctions.

The NEGF formalism for steady-state transport operates with two central quantities, the retarded ${\bf G}(E)$ and the lesser Green functions ${\bf G}^<(E)$, which describe the density of available quantum states and how electrons occupy those states, respectively~\cite{Haug2007}. Its application to  electronic transport is often combined~\cite{Datta1995} with the tight-binding (TB) Hamiltonian whose hopping parameters are fitted using more microscopic theory.\footnote{For example, in the case of either bulk graphene~\cite{Reich2002} or GNRs~\cite{Cresti2008} one has to employ TB Hamiltonian with up to third-nearest-neighbor hoppings in order to match the DFT-computed band structure.} This procedure is exemplified by the recent calculations~\cite{Markussen2009a} predicting $ZT \simeq 3$ for sufficiently long SiNW of 2 nm diameter with surface disorder, where electronic subsystem in silicon was described by sp$^3$d$^5$s$^*$ TB Hamiltonian with nearest-neighbor hoppings and ten orbitals per site.

However, such usage of NEGF formalism, where the device Hamiltonian is known from the outset, is not suitable for the description of realistic single-molecule nanojunctions where organic molecule consists of carbon and other atomic species, or in the case of GNR electrodes where different atoms or atomic groups are used to passivate dangling bonds along the edges~\cite{Cervantes-Sodi2008}. In such cases, {\em first-principles} input about atomistic and electronic structure is necessary in order to capture {\em charge transfer} between different atoms in equilibrium, geometrically-optimized atomic positions of the molecular bridge including molecule-electrode separation in equilibrium, and forces on atoms when they are perturbed out of equilibrium. For example, Ref.~\cite{Areshkin2010} shows how linear-response conductance of GNR-based devices is computed incorrectly if charge transfer between edge hydrogen and interior carbon atoms is not taken into account.

The state-of-the-art approach that can capture these effects, as long as the coupling between the molecule and the electrodes is strong enough to ensure transparent contact and diminish Coulomb blockade effects~\cite{Murphy2008,Cuniberti2005}, is NEGF-DFT. The DFT part of  this framework is employed using typical approximations (such as LDA, GGA, or B3LYP) for its exchange-correlation functional~\cite{Fiolhais2003}. The sophisticated computational algorithms~\cite{Areshkin2010,Cuniberti2005,Taylor2001,Brandbyge2002,Stokbro2008,Rungger2008} developed to implement the \mbox{NEGF-DFT} framework over the past decade can be encapsulated by the iterative self-consistent loop:
\begin{equation}\label{eq:scloop}
n^{\rm in}({\bf r}) \Rightarrow {\rm DFT} \rightarrow {\bf H}_{\rm KS}[n({\bf r})] \Rightarrow {\rm NEGF} \rightarrow n^{\rm out}({\bf r}).
\end{equation}
The loop starts from the initial input electron density $n^{\rm in}({\bf r})$ and then employs some standard DFT code~\cite{Fiolhais2003} typically in the basis set of finite-range orbitals for the valence electrons which allows for faster numerics and unambiguous partitioning of the system into the active region and the semi-infinite ideal electrodes. The DFT part of the calculation yields the single particle \mbox{Kohn-Sham} Hamiltonian
\begin{eqnarray}\label{eq:ks}
\hat{H}_{\rm KS}[n({\bf r})] & = & -\frac{\hbar^2\nabla^2}{2m}  + V^{\rm eff}({\bf r}), \\
V^{\rm eff}({\bf r}) & = & V_H({\bf r}) + V_{\rm xc}({\bf r}) + V_{\rm ext}({\bf r}).
\end{eqnarray}
Here $V^{\rm eff}({\bf r})$ is the DFT mean-field potential due to other electrons where $V_H({\bf r})$ is the Hartree, $V_{\rm xc}({\bf r})$ is the exchange-correlation and $V_{\rm ext}({\bf r})$ is the external potential contribution.  The inversion of $\hat{H}_{\rm KS}[n({\bf r})]$ yields the retarded Green function ${\bf G}(E)$ [see Eq.~(\ref{eq:gr}) below] whose integration over energy determines the density matrix via
\begin{equation}\label{eq:rho}
{\bm \rho}  =  \frac{1}{2\pi i} \int dE\, \mathbf{G}^<(E) = \int dE\, {\bf G}(E) [i f_L(E) {\bm \Gamma}_L(E) +  i f_R(E) {\bm \Gamma}_R(E)] {\bf G}^\dagger(E).
\end{equation}
Here the coherent transport regime (i.e., electron-phonon or electron-electron dephasing processes are absent) is assumed, so that $\mathbf{G}^<(E)$ can be expressed in terms of $\mathbf{G}(E)$. The matrix elements $n^{\rm out}(\bf r)=\langle {\bf r} | {\bm \rho} | {\bf r} \rangle$ are the new electron density  as the starting point of the next iteration. This procedure is repeated until the convergence criterion $||{\bm \rho}^{\rm out} - {\bm \rho}^{\rm in}|| < \delta$ is reached, where $\delta \ll 1$ is a tolerance parameter. The efficient computational algorithms for the construction of the density matrix in Eq.~(\ref{eq:rho})  for two-terminal device are discussed in Refs.~\cite{Areshkin2010,Brandbyge2002}, and the recently developed algorithms for ${\bm \rho}$ in  multiterminal devices (including multiterminal thermoelectrics~\cite{Saha2011}) can be found in Ref.~\cite{Saha2009a}.

The representation of the retarded Green function in the local orbital basis $\{ \phi_i \}$ requires to compute the inverse matrix
\begin{equation}\label{eq:gr}
{\bf G}(E) = [E{\bf S} - {\bf H} - {\bm \Sigma}_L(E) - {\bm \Sigma}_R(E)]^{-1},
\end{equation}
where the  Hamiltonian matrix ${\bf H}$ is composed of elements $H_{ij} = \langle \phi_i |\hat{H}_{\rm KS}| \phi_{j} \rangle$. The overlap matrix ${\bf S}$ has elements $S_{ij} = \langle \phi_i | \phi_j \rangle$. The non-Hermitian matrices ${\bm \Sigma}_{L,R}(E)$ represent the retarded self-energies due to the ``interaction'' with the left (L) and the right (R) electrodes~\cite{Haug2007,Datta1995}. These self-energies determine escape rates of electrons from the active region into the semi-infinite ideal electrodes, so that an open quantum system can be viewed as being described by the non-Hermitian Hamiltonian ${\bf H}_{\rm open}={\bf H}_{\rm KS}[n({\bf r})] + {\bm \Sigma}_L(E) + {\bm \Sigma}_R(E)$. The matrices ${\bm \Gamma}_{L,R}(E)=i[{\bm \Sigma}_{L,R}(E) - {\bm \Sigma}_{L,R}^\dagger(E)]$ account for the level broadening due to the coupling to the electrodes~\cite{Haug2007,Datta1995}.

The retarded Green function is required to post-process the result of the DFT loop and obtain the transport quantities. For example, the zero-bias electron transmission function in the elastic (no electron-electron~\cite{Bergfield2011} or electron-phonon~\cite{Frederiksen2007,Dash2011}  interactions) transport regime between the left and the right electrode is given by:
\begin{equation}\label{eq:telectron}
\mathcal{T}_{\rm el}(E) = {\rm Tr} \left\{ {\bm \Gamma}_R (E)  {\bf G}(E) {\bm \Gamma}_L (E)  {\bf G}^\dagger(E)  \right\}.
\end{equation}
Each electrode terminates at infinity into macroscopic reservoirs where electrons are assumed to be thermalized with their Fermi distribution function being $f(E-\mu_L)=f_L(E)$ or $f(E-\mu_R)=f_R(E)$ for the left and right reservoirs, respectively. The transmission function $\mathcal{T}_{\rm el}(E)$ in Eq.~(\ref{eq:telectron}) is then employed to compute the following auxiliary integrals~\cite{Esfarjani2006}
\begin{equation}\label{eq:kintegral}
K_n(\mu) = \frac{2}{h} \int\limits_{-\infty}^{\infty} dE\, \mathcal{T}_{\rm el}(E)  (E - \mu)^n \left(-\frac{\partial f(E,\mu)}{\partial E} \right),
\end{equation}
where \mbox{$f(E,\mu)=\{ 1 + \exp[(E-\mu)/k_BT] \}^{-1}$} is the Fermi distribution function at the chemical potential $\mu$ (which is the same for both reservoirs in the linear-response regime $\mu_L - \mu_R \rightarrow 0$). The knowledge of $K_n(\mu)$ finally yields all electronic quantities
\begin{eqnarray}
G & = & e^2K_0(\mu), \label{eq:conductance} \\
S & = & \frac{K_1(\mu)}{eTK_0(\mu)}, \label{eq:s} \\
\kappa_{\rm el} & = & \frac{K_2(\mu) - [K_1(\mu)]^2/K_0(\mu)}{T}. \label{eq:kappael}
\end{eqnarray}
that enter into the expression for $ZT$ in Eq.~\eqref{eq:zt}.

Our MT-NEGF-DFT code~\cite{Saha2010,Saha2009a}, employed to obtain results in Sec.~\ref{sec:electron} via the formulas discussed in this Section, utilizes  ultrasoft pseudopotentials and Perdew-Burke-Ernzerhof (PBE) parametrization of the generalized gradient approximation (GGA) for the exchange-correlation functional of DFT. The localized basis set for DFT calculations  is constructed from atom-centered orbitals (six per C atom and four per H atom with atomic radius 8.0 Bohr)  that are optimized variationally for the electrodes and the central molecule separately while their electronic structure is obtained concurrently.

We note that the well-known tendency of DFT-PBE to underestimate HOMO-LUMO ((HOMO-highest occupied molecular orbital; LUMO-lowest unoccupied molecular orbital) energy gap in molecules, and hence overestimate the conductance~\cite{Strange2011} does not influence the results presented in Sec.~\ref{sec:electron}. This is because for the nanojunctions studied here, the transmission around the Fermi level is not determined by the off-resonant tunneling through the HOMO-LUMO gap. Rather electrons propagate via resonant evanescent modes of the graphene electrodes. The energy and shape of such wavefunctions are determined by the electronic structure of the graphene nanoribbons and should be well described by DFT.

\section{NEGF-DFT methodology for the computation of phonon transport-determined thermoelectric quantities}\label{sec:negfph}

In contrast to rapid development~\cite{Haug2007,Datta1995} of NEGF-TB and NEGF-DFT-based quantum transport methods for electron propagation through nanostructures discussed in Sec.~\ref{sec:negfe}, comparable methods for phonon transport have emerged relatively slow~\cite{Wang2008c}. For example, classical molecular dynamics (MD) and the Boltzmann equation~\cite{McGaughey2004} are widely used traditional methods in phonon transport. However, the MD method~\cite{McGaughey2006} is not accurate below the Debye temperature, while the Boltzmann equation cannot be used in nanostructures without translational invariance. In both cases, the quantum effects become important.

Only recently, the NEGF formalism has been extended to study the quantum phononic transport~\cite{Wang2008c}. Nevertheless, among recent theoretical studies of single-molecule thermoelectric devices using first-principles quantum transport  frameworks~\cite{Pauly2008,Ke2009,Finch2009,Liu2009,Liu2009a,Liu2011a,Quek2011,Nozaki2010,Saha2011,Sergueev2011}, most have focused on computing the thermopower $S$, with only a few~\cite{Liu2009a,Nozaki2010,Saha2011,Sergueev2011} employing DFT to obtain forces on displaced atoms and then compute $\kappa_{\rm ph}$ from first-principles. The experimental data on the thermal conductance of single-molecule nanojunctions is even more scarce~\cite{Malen2010,Wang2006h}.

The phonon thermal conductance in the absence of phonon-phonon~\cite{Mingo2006} or electron-phonon~\cite{Frederiksen2007,Dash2011}  scattering is obtained from the phonon transmission function $\mathcal{T}_{\rm ph}(\omega)$ using the Landauer-type formula~\cite{Rego1998}:
\begin{equation}\label{eq:kappaphonon}
\kappa_{\rm ph} = \frac{\hbar^2}{2\pi k_B T^2} \int\limits_{0}^{\infty} d\omega\, \omega^2 \mathcal{T}_{\rm ph}(\omega) \frac{ e^{\hbar\omega/k_BT}}{(e^{\hbar\omega/k_BT}-1)^2}.
\end{equation}
The phonon transmission function $\mathcal{T}_{\rm ph} (\omega)$ in such elastic transport regime can be expressed in terms of NEGFs for the active region (molecule + portion of electrodes) attached to two semi-infinite electrodes
\begin{equation}\label{eq:transphonon}
\mathcal{T}_{\rm ph}(\omega) = {\rm Tr} \left\{ {\bm \Lambda}_R (\omega)  {\bf D}(\omega) {\bm \Lambda}_L (E)  {\bf D}^\dagger(\omega)  \right\},
\end{equation}
where the phonon GF is obtained in the same fashion as the electronic one in Eq.~\eqref{eq:gr} but with substitutions ${\bf H} \rightarrow {\bf K}$, $E{\bf S} \rightarrow \omega^2 {\bf M}$ and ${\bm \Sigma}_{L,R} \rightarrow {\bm \Pi}_{L,R}$:
\begin{equation}\label{eq:grph}
{\bf D}=[\omega^2 {\bf M} - {\bf K} - {\bm \Pi}_L - {\bm \Pi}_R]^{-1}.
\end{equation}
Here ${\bf K}$ is the force constant matrix, ${\bf M}$ is a diagonal matrix with the atomic masses, ${\bm \Pi}_{L,R}$ are the self-energies, and ${\bm \Lambda}_{L,R}(\omega)=i[{\bm \Lambda}_{L,R}(\omega) - {\bm \Lambda}_{L,R}^\dagger(\omega)]$.

We compute the force constant matrix ${\bf K}$ using GPAW~\cite{gpaw}, which is a real space electronic structure code based on the projector augmented wave method~\cite{Enkovaara2010}. The electronic wavefunctions are expanded in atomic orbitals with a single-zeta polarized basis set, and PBE exchange-correlation functional is used. The whole active region, which includes 27 layers of ZGNR electrodes, is first relaxed to a maximum force of $0.01 \ \mathrm{eV/\AA}$  per atom.

Subsequently, we displace each atom $I$ by $Q_{I\alpha}$ in the direction $\alpha=\{x,y,z\}$ to get the forces $F_{J\beta}(Q_{I\alpha})$ on atom $J\neq I$ in direction $\beta$. The elements of ${\bf K}$-matrix are then computed from finite differences
\begin{equation}\label{eq:kmatrix}
K_{I\alpha,J\beta} = \frac{F_{J\beta}(Q_{I\alpha})-F_{J\beta}(-Q_{I\alpha})}{2Q_{I\alpha}}.
\end{equation}
The intra-atomic elements are calculated by imposing momentum conservation, such that $K_{I\alpha,I\beta} = -\Sigma_{J\neq I}K_{I\alpha,J\beta}$.

The methodology described in this Section does not take into account resistive umklapp phonon-phonon scattering which plays an important role in interpretation of experiments on room temperature thermal conductivity of large-area graphene~\cite{Ghosh2009,Seol2010,Chen2011}. However, this effect,  which is easy to describe using the Boltzmann equation but is very expensive computationally within NEGF formalism~\cite{Mingo2006}, should not play an important role in devices depicted in Fig.~\ref{fig:fig1} because the mean-free path in graphene is \mbox{$\ell \simeq 677$ nm} at room temperature~\cite{Aksamija2011}. That is, the active region  of the single-molecule nanojunctions in Fig.~\ref{fig:fig1} or the width of their GNR electrodes is much smaller than $\ell$ at all temperatures below the ambient one which are the focus of our study.

\section{Electronic transmission and thermopower in single-molecule nanojunctions with GNR electrodes}\label{sec:electron}

The recent fabrication of GNRs with ultrasmooth edges~\cite{Cai2010,Jia2009,Tao2011} has opened new avenues for highly controllable molecular junctions with a well-defined molecule-electrode contact characterized by  high transparency, strong directionality and reproducibility. This is due to the fact that strong molecule-GNR \mbox{$\pi$-$\pi$} coupling makes possible  formation of a continuous \mbox{$\pi$-bonded} network across GNR and orbitals of $\pi$-conjugated organic molecules~\cite{Ke2007}. Although GNRs with zigzag~\cite{Jia2009} or chiral edges~\cite{Tao2011} are insulating at very low temperatures due to one-dimensional spin-polarized edge states coupled across the width of the nanoribbon, such unusual magnetic ordering and the corresponding band gap is easily destroyed above $\gtrsim 10$ K~\cite{Yazyev2008,Kunstmann2011} so that they can be considered as candidates for metallic electrodes. In fact, the experimental pursuit of graphene-based single-molecule nanojunctions, where a $\pi$-conjugated organic molecule is inserted into the nanogap formed by feedback controlled electroburning of few-layer graphene, has commenced very recently~\cite{Prins2011}.

\begin{figure}
\begin{center}
\includegraphics[scale=0.45,angle=0]{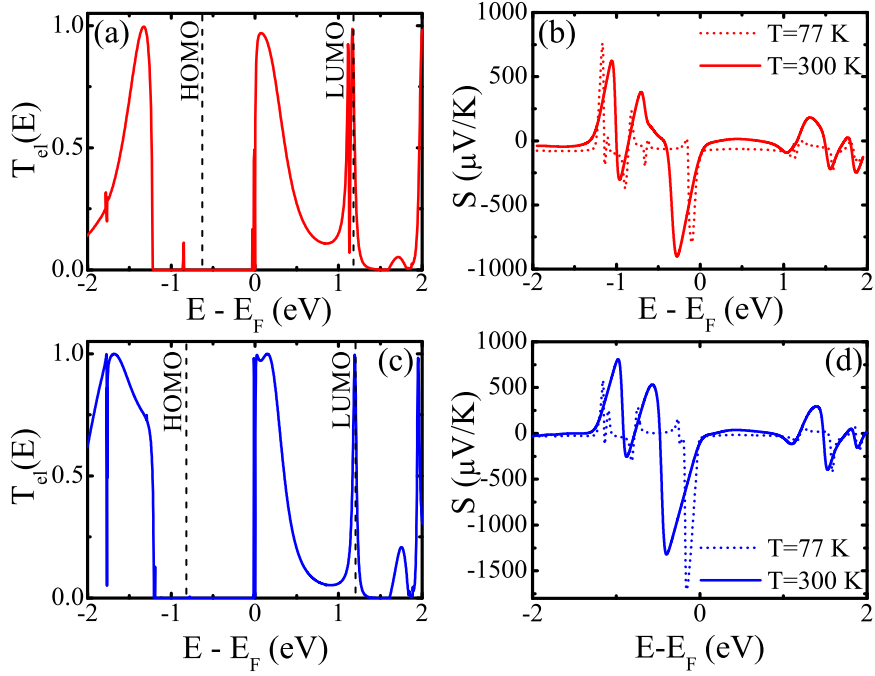}
\end{center}
\caption{(a) Zero-bias electronic transmission $\mathcal{T}_{\rm el}(E)$ for ZGNR$|$18-annulene$|$ZGNR junction; (b) thermopower at two different temperatures for ZGNR$|$18-annulene$|$ZGNR junction; (c) zero-bias electronic transmission for ZGNR$|$C10$|$ZGNR junction; and (d) thermopower at two different temperatures for ZGNR$|$C10$|$ZGNR junction. The vertical dashed lines in panels (a) and (c) mark the position of HOMO and LUMO levels of 18-annulene or C10 molecules after they are attached to the electrodes, as extracted from the projected density of states for carbon atoms in the center of the junction.}
\label{fig:fig2}
\end{figure}

Unlike the metallic carbon nanotubes (CNTs), which were employed in earlier experiments~\cite{Guo2006} as electrodes of single-molecule nanojunctions in order to generate \mbox{$\pi$-bonded} network~\cite{Ke2007}, GNRs have planar structure appropriate for aligning and patterning. The early experiments~\cite{Guo2006} on CNT$|$molecule$|$CNT heterojunctions  have measured surprisingly small conductances for a variety of  sandwiched molecules. The first-principles analysis of different setups reveals that this is due to significant twisting forces when molecule is connected to CNT via 6-membered rings~\cite{Ke2007}. Therefore, to keep nearly parallel and in-plane configuration (hydrogen atoms of \mbox{18-annulene} slightly deviate from the molecular plane) of our ZGNR$|$18-annulene$|$ZGNR junction, we use a 5-membered ring~\cite{Ke2007}  to chemically bond ZGNR to 18-annulene or to C10 molecule, as illustrated in Fig.~\ref{fig:fig1}.

The high contact transparency makes it possible for evanescent wavefunctions from the two ZGNR electrodes to tunnel into the molecular region and meet in the middle of it (when the molecule is short enough~\cite{Ke2007}). This is a counterpart of  the well-known metal induced gap states  in metal-semiconductor Schottky junctions. Such effect can induce a large peak (i.e., a resonance) in the electronic transmission function near $E_F$,  despite the HOMO-LUMO  energy gap of the isolated molecule. This  phenomenology is confirmed by Figs.~\ref{fig:fig2}(a) and ~\ref{fig:fig2}(c) showing the zero-bias electronic transmission $\mathcal{T}_{\rm el}(E)$  where the peak around $E-E_F=0$ is conspicuous for {\em both} single-molecule nanojunctions illustrated in Fig.~\ref{fig:fig1}. The peak is located far away from HOMO and LUMO levels of the molecules which are marked in Figs.~\ref{fig:fig2}(a) and (c).

The carbon atoms of a ring-shaped \mbox{18-annulene} molecule can be connected to ZGNR electrodes in configurations whose Feynman paths for electrons traveling around the ring generate either constructive or destructive quantum interference effects imprinted on the conductance~\cite{Markussen2010a,Markussen2011}. For example, a $\pi$-electron at $E_F$ entering the molecule in setup shown in Fig.~\ref{fig:fig1} has wavelength $k_F/2d$ ($d$ is the spacing between carbon atoms within the molecule), so that for the two simplest Feynman paths of length $9d$ (upper half of the ring) and $9d$ (lower half of the ring) the phase difference is 0 and constructive interference occurs.  Note that the destructive quantum interference~\cite{Markussen2010a,Markussen2011} would form an additional dip~\cite{Saha2010} (i.e., an antiresonance) within the main transmission peak around $E-E_F=0$ in Fig.~\ref{fig:fig2}(a). The effect of such antiresonance on the thermopower $S$ for gold$|$18-annulene$|$gold junctions has been analyzed in Ref.~\cite{Bergfield2009} as a possible sensitive tool to confirm the effects of quantum coherence, even at room temperature, on transport through single-molecule nanojunctions.

Additionally, the suppression of the hole-like transmission, $\mathcal{T}_{\rm el}(E) \rightarrow 0$ for $-1.0 \ \mathrm{eV} \lesssim E-E_F < 0$, avoids  unfavorable compensation~\cite{Nozaki2010} of hole-like and electron-like contributions to the thermopower. This is due to the symmetry of transverse propagating eigenstates in the semi-infinite ZGNR electrodes which we elaborate in more detail in Sec.~\ref{sec:symmetry}. These features in the electronic transmission function yield the thermopower $S$ in Figs.~\ref{fig:fig2}(b) and ~\ref{fig:fig2}(d) whose maximum value, which is slightly away from $E-E_F=0$ at room temperature, is an {\em order of magnitude larger} than the one measured in large-area graphene~\cite{Zuev2009} or in organic molecules sandwiched by gold electrodes~\cite{Reddy2007}.

Comparing Figs.~\ref{fig:fig2}(a) and ~\ref{fig:fig2}(b) for ZGNR$|$18-annulene$|$ZGNR junction with Figs.~\ref{fig:fig2}(c) and ~\ref{fig:fig2}(d) for ZGNR$|$C10$|$ZGNR junction shows great similarity of the transmission function around the Fermi energy $E-E_F=0$. This is due to the transport through overlapping evanescent modes, so that the same features in the electronic transmission function and thermopower will persist for any {\em sufficiently short}~\cite{Ke2007,Ke2008} organic molecule as long as its HOMO and LUMO levels [whose position is marked in Figs.~\ref{fig:fig2}(a) and (c)] are far away from the Fermi energy of the electrodes. Since these features in the transmission function are governed by evanescent and propagating wavefunctions originating in metallic ZGNR electrodes, they are {\em impervious} to the usual poor estimates of energy gaps and molecular level position in DFT and, therefore, do not require more accurate but computationally much more expensive NEGF-GW approach~\cite{Strange2011}.

\section{Suppression of hole-like transmission by the symmetry of transverse propagating states in ZGNR electrodes}\label{sec:symmetry}

\begin{figure}%[htb!]
\begin{center}
\includegraphics[width=\columnwidth]{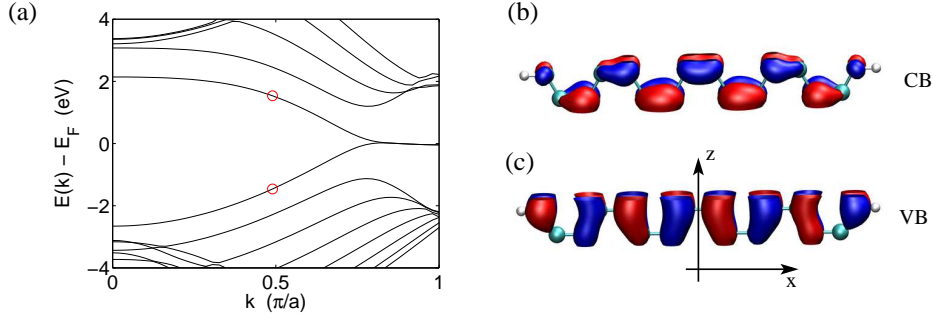}
\end{center}
    \caption{(a) The band structure of 8-ZGNR. Imaginary part of (b) conduction band and (c) valence band Bloch eigenstates evaluated at $k= \pi/2a$, as indicated with circles in panel (a). The conduction band state is symmetric with respect to mirroring in the $z$-axis, while the valence band state is antisymmetric.}
\label{zgnr-band-fig}
\end{figure}

The evanescent mode-induced transmission resonance can also be generated in setups where CNT electrodes are attached to a short organic  molecule\footnote{For an example of the peak in $T_{\rm el}(E)$ induced by the overlap of evanescent wavefunctions originating from two CNT electrodes sandwiching 18-annulene molecule see Supplemental Material of Ref.~\cite{Ke2007}.} via transparent contacts~\cite{Ke2007,Ke2008} However, besides the effect of the same type in our ZGNR$|$18-annulene$|$ZGNR and ZGNR$|$C10$|$ZGNR junctions, their power factor $S^2G$ is further enhanced by completely different transport properties of the ZGNR conduction band (CB) Bloch eigenstates (exhibiting high transmission) and the valence band (VB) Bloch eigenstates (exhibiting low transmission). The origin of this asymmetry can be explained by analyzing the spatial symmetry of the CB and VB propagating eigenstates. Figure~\ref{zgnr-band-fig}(a) shows the DFT-calculated band structure of infinite homogeneous ZGNR. The right part of the figure shows isosurface plots of the Bloch eigenstates (imaginary part only) in the CB [Fig.~\ref{zgnr-band-fig}(b)] and VB [Fig.~\ref{zgnr-band-fig}(c)] evaluated at $k=\pi/2a$ ($a$ is the lattice constant of graphene).

The CB and VB states have different symmetry properties---while the CB state is symmetric with respect to mirroring in the $z$-axis, the VB state is antisymmetric. Recall that molecules in Fig.~\ref{fig:fig1} are coupled to the electrodes via a pentagon in the middle of the transverse edge of the nanoribbon, and thus form a two-atom linear carbon chain between the ZGNR and the rest of the molecule. The state in this linear chain will certainly be symmetric so that they do not couple with the antisymmetric states in VB, while a large coupling to the symmetric states in CB is expected.

\begin{figure}%[htb!]
\begin{center}
\includegraphics[width=0.8\columnwidth]{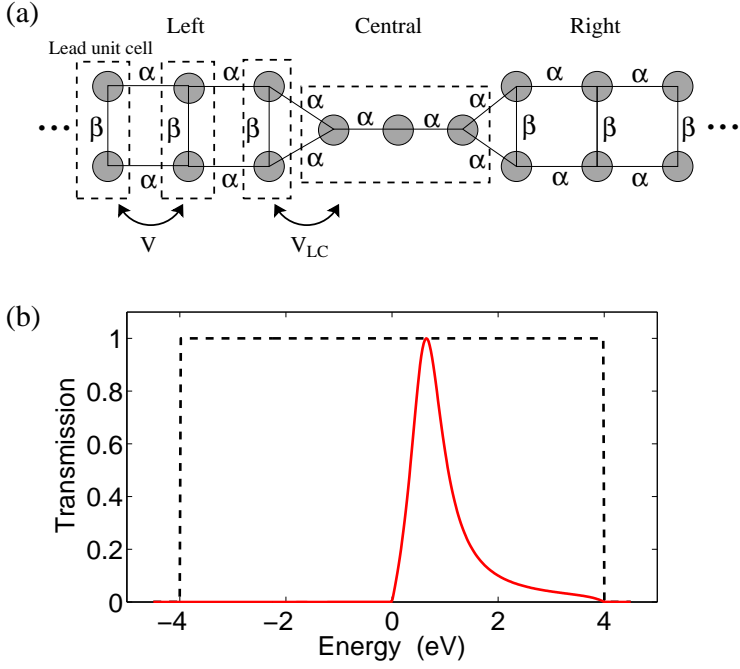}
\end{center}
    \caption{(a) Simple model of a two-terminal junction consisting of two semi-infinite double tight-binding chains, which act as the left and the right lead attached to a single tight-binding chain as the central region. The transmission function for this device is shown in panel (b) as the solid red line, while the transmission function for the infinite double-chain lead is shown in dashed black for comparison.}
\label{double-single-double-fig}
\end{figure}

Although the above explanation of the shape of transmission function $\mathcal{T}_{\rm el}(E)$ around $E-E_F=0$ in terms of the spatial symmetry properties of CB and VB Bloch eigenstates is rather intuitive, it may be instructive to show in more detail the connection between the eigenstates of the electrodes and the NEGF formalism which often makes no explicit reference to such eigenstates~\cite{Velev2004}. For this purpose, we consider a simple model depicted in Fig.~\ref{double-single-double-fig}(a) which shares many features with single-molecule nanojunctions in Fig.~\ref{fig:fig1}. The model consists of two semi-infinite  double tight-binding chains acting as leads that are attached to a single tight-binding chain in the center of the device. We assume a single $s$-orbital on each site with on-site energy $\varepsilon_0=0$ eV. The hopping parameters are $\beta=2$ eV for inter-chain hoppings and $\alpha=1$ eV for all other hoppings.

Thus, the Hamiltonian of the lead unit cell marked by rectangles in Fig.~\ref{double-single-double-fig}(a) is given by
\begin{eqnarray}
\mathbf{H}_0= \left(
\begin{array}{cc}
	\varepsilon_0 & \beta \\
	\beta & \varepsilon_0
\end{array} \right),
\end{eqnarray}
and the coupling matrix between two such unit cells is
\begin{eqnarray}
\mathbf{V} =\left(
\begin{array}{cc}
	\alpha & 0 \\
	0 & \alpha
\end{array} \right)
\end{eqnarray}
Figure~\ref{double-single-double-fig}(b) plots the transmission function of the junction (solid red) and of the infinite electrodes (dashed black). Similar to $\mathcal{T}_{\rm el}(E)$ plotted in Fig.~\ref{fig:fig2}(a) and ~\ref{fig:fig2}(c), we find a complete blocking of the transmission function for  VB states, while the CB states are highly transmitting. This asymmetry between quantum transport through VB and CB transverse propagating states may again be explained by their symmetry properties.

In analogy with ZGNR, the VB Bloch eigenstate is antisymmetric while the CB Bloch eigenstate is symmetric. The Bloch eigenstates can be written as
\begin{eqnarray}
\psi_{VB,CB}(k) & = & \phi_{VB,CB}e^{-ikz}, \\
\phi_{VB,CB} & = & \frac{1}{\sqrt{2}} \left(\begin{array}{c}
1 \\
a
\end{array}\right),
\end{eqnarray}
where $a=-1$ for VB and $a=+1$ for CB. We now recall that the lead self-energy can be expressed as~\cite{Lopez-Sancho1984}
\begin{eqnarray}
{\bm \Sigma}_L = -\mathbf{V} \cdot \mathbf{T},
\end{eqnarray}
using the transfer matrix $\mathbf{T}$. For our purpose it is sufficient to consider only the imaginary part of the self-energy because the asymmetry is determined by ${\bm \Gamma_L}=-2{\rm Im}\, [{\bm \Sigma_L}]$. The transfer matrix can be calculated from the {\em complex band structure}~\cite{Rungger2008} including both propagating and evanescent states. The imaginary part of the transfer matrix is determined solely by the transverse propagating states~\cite{Rungger2008}:
\begin{eqnarray}
{\rm Im}\, [T(E)] = \sum_{n}\phi_n{\rm Im}(e^{ik_n})\phi_n^\dagger,
\end{eqnarray}
where the sum runs only over propagating states whose transverse part $\phi_n$ is purely real. The surface Green function~\cite{Velev2004} in the lead-unit cell next to the central single chain is related to the self-energy by $\mathbf{g_L}=\mathbf{V}^{-1}{\bm \Sigma}_L\mathbf{V}^{-1}$.
It now follows that the surface Green function has the structure
\begin{eqnarray}
{\rm Im}\, [\mathbf{g}_L] \propto \left(
\begin{array}{cc}
	1 & a \\
	a & a^2
\end{array} \right) =
\left(
\begin{array}{cc}
	1 & a \\
	a & 1
\end{array} \right). \label{eq:gL}
\end{eqnarray}

The self-energy on the single chain due to the left lead is obtained as ${\bm \Sigma}_{L}=\mathbf{V}_{LC}^\dagger\mathbf{g}_L\mathbf{V}_{LC}$, where $\mathbf{V}_{LC}$ is the coupling matrix between the left lead and the central single-chain (there is a similar contribution from the right lead as discussed in Sec.~\ref{sec:negfe}). Since only the first site in the chain is connected to the left lead, ${\bm \Sigma}_{L}$ has only one non-zero element  $({\bm \Sigma}_{L})_{11} \neq 0$ in the upper left corner. From Eq.~\eqref{eq:gL} we get that ${\rm Im}\, [({\bm \Sigma}_L)_{11}] \propto (1+a)$, such that ${\bm \Gamma}_{L}=-2 {\rm Im}\, [{\bm \Sigma}_{L}] = \mathbf{0}$ for energies in VB, while   $({\bm \Gamma}_{L})_{11}$ is non-zero for energies in CB. Therefore, since the transmission function of the junction is given by Eq.~\eqref{eq:telectron}, it follows that it must be zero for energies in the VB.

The analysis presented in this Section demonstrates that the vanishing transmission function [Fig.~\ref{double-single-double-fig}(b)] at energies within the VB for the simple model junction, as well as vanishing hole-like transmission function [Figs.~\ref{fig:fig2}(a) and (c)] in the energy range below $E-E_F=0$ in realistic ZGNR$|$18-Annulene$|$ZGNR or ZGNR$|$C10$|$ZGNR junctions, can be explained by the symmetry properties of the transverse propagating states in the semi-infinite electrodes. This demonstrate the generality of our ZGNR$|$molecule$|$ZGNR device concept and explains why different molecules can have very similar transmission functions, as exemplified by Figs.~\ref{fig:fig2}(a) and ~\ref{fig:fig2}(c). On the other hand, it also illustrates a weakness in the setup because the central molecule must be coupled to one of the ZGNRs exactly in the middle if its transverse  edge. Nevertheless, these findings open up further possibilities to search for other single-molecule nanojunctions which can exploit the symmetry of the Bloch states in their electrodes in order to optimize the thermopower.
\section{Phonon transmission and thermal conductance in single-molecule nanojunctions with GNR electrodes}\label{sec:phonon}

\begin{figure}
\centerline{\includegraphics[scale=0.29,angle=0]{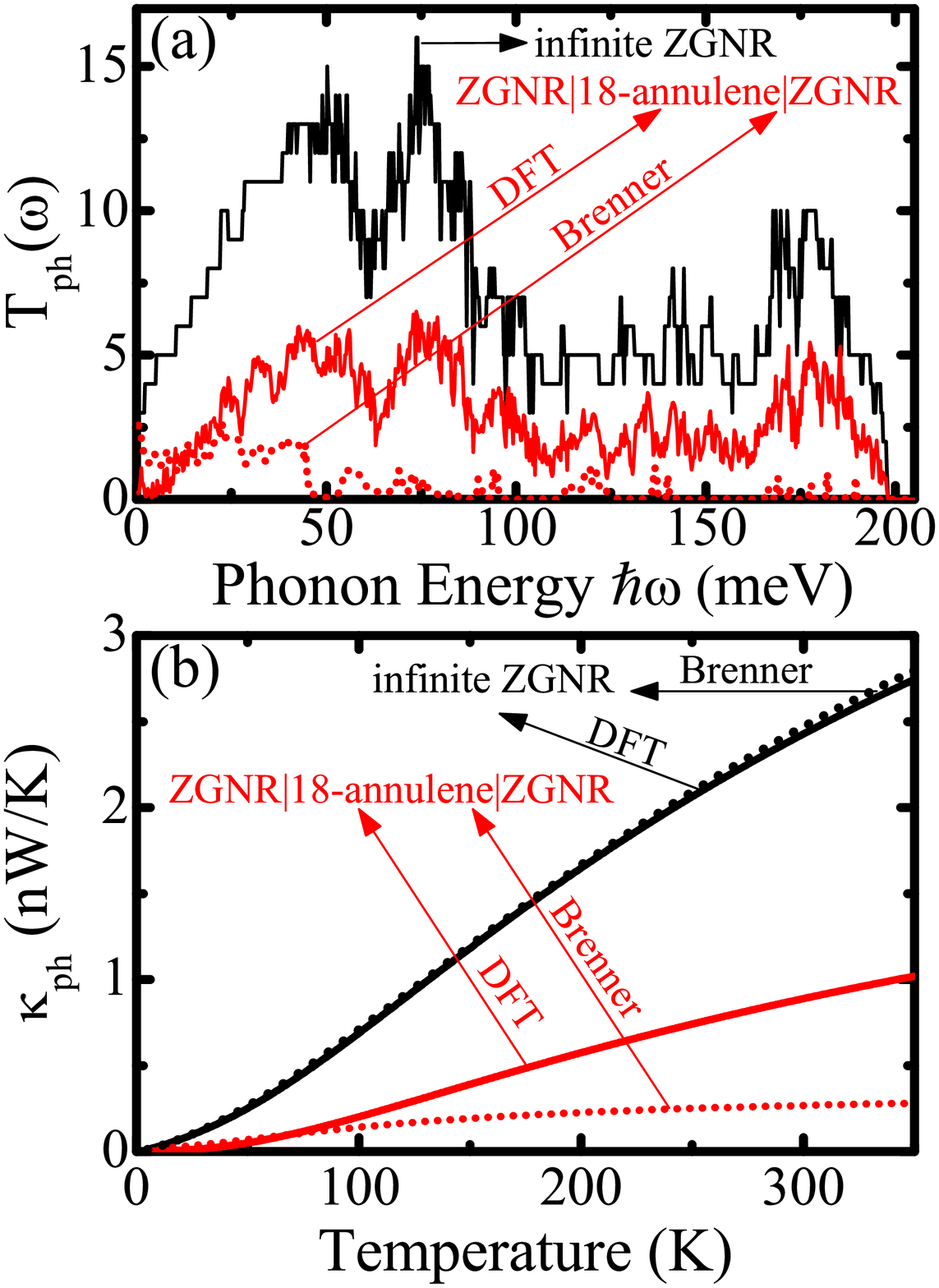} \includegraphics[scale=0.29,angle=0]{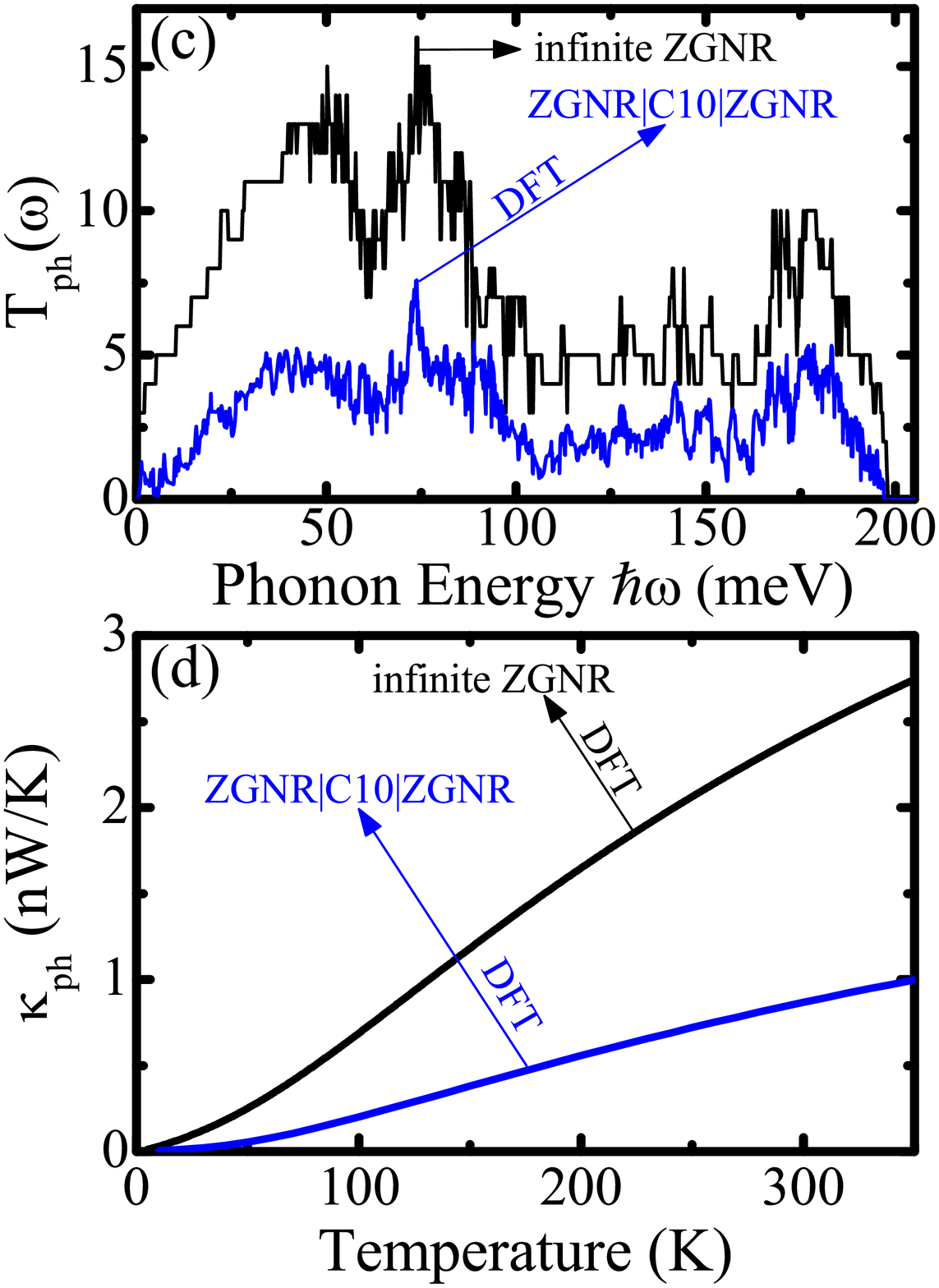}}
\caption{(a) The phonon transmission function $\mathcal{T}_{\rm ph}(\omega)$ and (b) the corresponding thermal conductance $\kappa_{\rm ph}$ for an
infinite 8-ZGNR and ZGNR$|$18-annulene$|$ZGNR single-molecule nanojunction shown in Fig.~\ref{fig:fig1}(a). (c) The phonon transmission function and
(d) the corresponding thermal conductance $\kappa_{\rm ph}$ for an infinite 8-ZGNR and ZGNR$|$C10$|$ZGNR single-molecule nanojunction shown in
Fig.~\ref{fig:fig1}(b). Panels (a) and (b) also compare results obtained using NEGF coupled to DFT (via GPAW code~\cite{gpaw}) vs. NEGF coupled to
Brenner EIP (via GULP code~\cite{gulp}).}
\label{fig:fig3}
\end{figure}

Figure~\ref{fig:fig3} shows the phonon transmission function $\mathcal{T}_{\rm ph}(\omega)$ and the corresponding phonon thermal conductance $\kappa_{\rm ph}$ for ZGNR$|$18-annulene$|$ZGNR and ZGNR$|$C10$|$ZGNR single-molecule nanojunctions computed via the first-principles formalism delineated in Sec.~\ref{sec:negfph}. Note that we use the term ``phonon'' here freely to refer to any quantized vibrations in the active region of the junction. To understand how the mismatch in vibrational properties of semi-infinite ZGNR electrodes and the molecule impedes phonon transport across the junction, we also plot $\mathcal{T}_{\rm ph}(\omega)$ [in Figs.~\ref{fig:fig3}(a) and ~\ref{fig:fig3}(c)] and $\kappa_{\rm ph}$ [in Figs.~\ref{fig:fig3}(b) and ~\ref{fig:fig3}(d)] for an infinite homogeneous 8-ZGNR.

The phonon transmission function for infinite homogeneous 8-ZGNR consists of quantized steps, as expected for purely ballistic transport of phonons described in harmonic approximation~\cite{Rego1998}. The suppression of  phonon transmission by the presence of the molecule generates three times smaller $\kappa_{\rm ph}$ at room temperature when compared  to the thermal conductance of an infinite 8-ZGNR, as shown in Figs.~\ref{fig:fig3}(b) and ~\ref{fig:fig3}(d).

\subsection{Comparison of DFT and semi-empirical methods for the computation of phonon thermal conductance}\label{sec:brenner}

The force constant matrix ${\bf K}$, discussed in Sec.~\ref{sec:negfph} as an input for NEGF calculations, can also be obtained using empirical interatomic potentials. These is widely used methodology to treat elastic phonon quantum transport in clear or disordered nanowires~\cite{Markussen2009a,Zimmermann2008,Sevincli2010} since it is computationally much less expensive that our first-principles approach discussed in Sections~\ref{sec:negfph} and ~\ref{sec:phonon}. One of the standard choices for hydrocarbon systems is the so-called Brenner EIP~\cite{Brenner1990}, which is often applied to study lattice dynamics and phonon thermal transport in carbon nanotubes and graphene~\cite{Lindsay2010}.

The Brenner EIPs are short range, so they cannot accurately fit the graphene dispersion over the entire Brillouin zone (BZ).  However, the thermal transport depends more sensitively on the accuracy of acoustic phonon frequencies near the zone center where the longitudinal- and transverse-acoustic (LA and TA) velocities and the quadratic curvature of the out-of-plane acoustic (ZA) branch are determined. Conversely, only weak thermal excitation of the optical phonons and acoustic phonons near the BZ boundary occurs around room temperature because of the large Debye temperature ($\sim 2000$ K)
of graphene.

The basic steps of NEGF-Brenner-EIP methodology are: initially relax the device geometry $\Rightarrow$ the force constant between atom $I$ in direction $\alpha={x,y,x}$ and atom $J$ in direction $\beta$ is calculated using analytical derivatives, $K_{I\alpha,J\beta} = \partial U/(\partial R_{I\alpha}\partial R_{J\beta})$, where $U$ is the total energy $\Rightarrow$ compute retarded GF in Eq.~\eqref{eq:grph} and phonon transmission
function using  Eq.~\eqref{eq:transphonon}. Here we employ Brenner EIP as implemented in the GULP code~\cite{gulp,Gale1997}.

\begin{figure}
\begin{center}
\includegraphics[scale=0.45,angle=0]{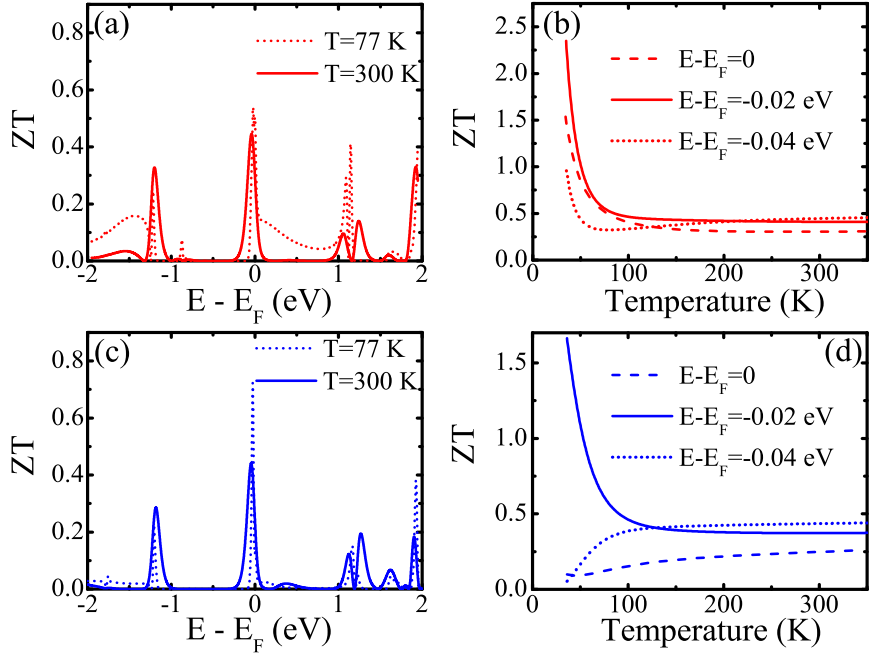}
\end{center}
\caption{The thermoelectric figure of merit $ZT$ for the two single-molecule nanojunctions shown in Fig.~\ref{fig:fig1}: (a) $ZT$ vs. energy at two different temperatures for ZGNR$|$18-annulene$|$ZGNR junction; (b) $ZT$ vs. temperature at three different energies for ZGNR$|$18-annulene$|$ZGNR junction; (c) $ZT$ vs. energy at two different temperatures for ZGNR$|$C10$|$ZGNR junction; and (d) $ZT$ vs. temperature at three different energies for ZGNR$|$C10$|$ZGNR junction.}
\label{fig:fig4}
\end{figure}

To contrast this methodology with the full NEGF-DFT treatment of the phonon thermal transport  in our single-molecule nanojunctions, we add NEGF-Brenner-EIP results for $\mathcal{T}_{\rm ph}(\omega)$ and  $\kappa_{\rm ph}$ for the case of ZGNR$|$18-annulene$|$ZGNR junction and infinite homogeneous 8-ZGNR in Figs.~\ref{fig:fig3}(a) and ~\ref{fig:fig3}(b). Although NEGF-Brenner-EIP calculations differs very little from NEGF-DFT in the case of infinite homogeneous ZGNR, the NEGF-Brenner-EIP methodology yields significantly lower $\mathcal{T}_{\rm ph}(\omega)$ and  $\kappa_{\rm ph}$ for ZGNR$|$18-annulene$|$ZGNR junction. Thus, using the Brenner-EIP to describe phonon transport would lead to an overestimated $ZT$.

\section{Thermoelectric figure of merit of single-molecule nanojunctions with GNR electrodes}\label{sec:zt}

Combining the first-principles results in Sec.~\ref{sec:electron} for electron transport and in Sec.~\ref{sec:phonon} for phonon transport allows us to obtain the thermoelectric figure of merit $ZT$. The results shown in Fig.~\ref{fig:fig4} demonstrate $ZT \sim 0.5$ for both junctions at room temperature. For comparison, we note that the recent proposal~\cite{Nozaki2010} for the single-molecule thermoelectric devices with sophisticated combination of local chemical tuning of the molecular states and usage of semiconducting electrodes has predicted much smaller $ZT \sim 0.1$ at room temperature.

In addition, $0.5<ZT<2.5$ at $E-E_F=-0.02$ eV (which can be set by the backgate electrode covering the whole device~\cite{Zuev2009}) in Fig.~\ref{fig:fig4} within the temperature range \mbox{$T = $ 30--77 K} is much larger than the value achieved in conventional low-temperature bulk thermoelectric materials~\cite{Snyder2008}. It is also larger than $ZT \simeq 1$ found in recent studies of low-temperatures ($T<100$ K) nanoscale thermoelectrics, such as the Kondo insulator nanowires~\cite{Zhang2011}. Thus, our single-molecule nanojunctions could be suitable for high-performance thermoelectric cooling at low temperatures.

\section{Summary and future prospects} \label{sec:conclusion}

We provided an overview of NEGF-DFT methodology to treat {\em independent} electron and phonon transport in nanoscale thermoelectric devices, where we focused on the nascent subfield of single-molecule nanojunctions. The examples of nanojunctions we considered consist of a $\pi$-conjugated organic molecule attached to metallic ZGNR electrodes via strong covalent bond which creates continuous \mbox{$\pi$-bonded} network across the device. Although devices in Fig.~\ref{fig:fig1} look futuristic at first sight, presently available nanofabrication technologies have already demonstrated feasibility of single-molecule nanojunctions with graphene electrodes~\cite{Prins2011}. The highly  transparent GNR$|$molecule contact allows for evanescent wavefunctions to penetrate from the ZGNR electrodes into the HOMO-LUMO gap of sufficiently short molecule, thereby generating a transmission resonance as a favorable ingredient to optimize the electronic thermopower. While such resonance could be achieved by using other types of carbon-based metallic electrodes such as metallic CNTs, the usage of ZGNRs also brings peculiar spatial symmetry properties of transverse propagating states which block transmission of electrons in a range of energies below the Fermi energy of the device. This increases the thermopower where, otherwise, contributions from hole-like and electron-like transmission appear with different signs in Eq.~\eqref{eq:s} and thus partially cancel each other. Finally, the mismatch in vibrational properties of the semi-infinite ZGNR electrode and the molecule acts to reduce the phonon thermal conductance across the junction by a factor of about three when compared to infinite homogeneous ZGNRs. The combination of these three features leads  to $ZT \sim 0.5$  at room temperature and $0.5<ZT <2.5$ below liquid nitrogen temperature for the junctions in Fig.~\ref{fig:fig1}. These values of $ZT$ turn out to be much higher than those found in other recent first-principles studies of single-molecule thermoelectric devices~\cite{Liu2009a,Nozaki2010,Sergueev2011}.

While the thermoelectric figure of merit $ZT$ we obtained at room temperature is still low, we anticipate that much higher $ZT$ could be achieved by {\em testing different types of molecules} to reduce $\kappa_{\rm ph}$ further. Essentially, the single-molecule nanojunctions proposed in Fig.~\ref{fig:fig1} should be viewed as representatives of a new class of nanoscale thermoelectric devices where the power factor $S^2G$ is already optimized by the usage of GNR electrodes that generate molecular-level-independent transmission resonance and where the spatial symmetry of transverse propagating eigenstates in GNR electrodes  lifts the compensation of hole-like and electron-like contributions to $S$.

\subsection{Future prospects}\label{sec:future}

We also offer a brief survey of possible future directions and related challenges. Our discussion of quantum electronic and phononic transport in realistic single-molecule nanojunctions via NEGF-DFT formalism has been confined to {\em non-interacting} electrons and phonons which propagate independently of each other. Since single-molecule nanojunctions depicted in Fig.~\ref{fig:fig1} have highly transparent contacts due to strong molecule-electrode coupling, we have not considered Coulomb blockade effects~\cite{Bergfield2011,Leijnse2010,Murphy2008} that would become relevant in molecules weakly coupled to the electrodes. Their understanding (such as simultaneous treatment of Coulomb blockade and coherent tunneling transport regimes~\cite{Bergfield2011}) to correctly capture thermoelectric properties in the presence of many-body interactions is in its infancy.

Although the electron-phonon interaction is the dominant mechanism of inelastic scattering in single-molecule nanojunctions~\cite{Mitra2004,Galperin2007,Frederiksen2007,Dash2011}, whose effect on electronic current~\cite{Mitra2004,Galperin2007,Dash2011,Horsfield2006} and its noise~\cite{Haupt2010} has attracted considerable attention, the study of electron and phonon quantum transport coupled by such interaction is {\em extremely rare}. Also, while the techniques have been developed to take into account anharmonicity into NEGF calculations of phonon transport through single-molecule nanojunctions~\cite{Mingo2006}, the treatment of phonon-phonon scattering via such fully quantum-mechanical approach combined with first-principles methods to capture atomistic structure of realistic junctions is lacking.

For example, there is only a handful of recent papers~\cite{Sergueev2011,Galperin2008,Lu2007a,Hsu2011,Asai2008,Jiang2011} where this problem of coupled electron and phonon transport has been tackled using NEGF applied to simple phenomenological models~\cite{Galperin2008,Lu2007a,Hsu2011,Asai2008,Jiang2011}, or NEGF-DFT with simplifications in considering interacting electrons and noninteracting phonons~\cite{Sergueev2011}. Nevertheless, these attempts hint~\cite{Sergueev2011,Galperin2008} at possibly strong effect  of electron-phonon interaction on the thermopower $S$, as well as that  $ZT$ of single-molecule nanojunctions can be enhanced by them. In fact, electron-phonon interaction is largely an unexplored parameter in the quest for efficient thermoelectric materials, as highlighted by the recent experiments on superlattices~\cite{Choi2010}.

The main technical issue in evaluating even the lowest order Feynman diagrams (Hartree and Fock for electrons~\cite{Frederiksen2007,Dash2011} and polarization bubble for phonons~\cite{Jiang2011}) for the retarded and lesser nonequilibrium self-energies~\cite{Haug2007} is integration in energy which has to be done repeatedly in order to achieve self-consistency that ensures conservation of charge and heat currents~\cite{Lu2007a}. The successful solution to this extremely computationally intensive~\cite{Frederiksen2007} problem will open a pathway to understand phonon drag (interchange of momentum between acoustic phonons and electrons that generates additional contribution to the thermopower which has thus far been studied only via coupled electron-phonon Boltzmann equations~\cite{Scarola2002}) and electron drag (where phonons are dragged by electrons from low $T$ region into high $T$ region~\cite{Jiang2011}) effects in realistic single-molecule thermoelectric devices using first-principles quantum transport simulations.

% For two-column wide figures use
%\begin{figure*}
% Use the relevant command to insert your figure file.
% For example, with the graphicx package use
%  \includegraphics[width=0.75\textwidth]{example.eps}
% figure caption is below the figure
%caption{Please write your figure caption here}
%\label{fig:2}       % Give a unique label
%\end{figure*}
%

\begin{acknowledgements}
We thank K. Esfarjani and M. Paulsson for illuminating discussions. Financial support under DOE Grant No. DE-FG02-07ER46374 (K. K. S. and B. K. N.) and FTP Grant No. 274-08-0408 (T. M. and K. S. T.) is gratefully acknowledged. The supercomputing time was provided in part by the NSF through TeraGrid resource TACC Ranger under Grant No. TG-DMR100002 and NSF Grant No. CNS-0958512.
\end{acknowledgements}

% BibTeX users please use one of
%\bibliographystyle{spbasic}      % basic style, author-year citations
%\bibliographystyle{spmpsci}      % mathematics and physical sciences
%\bibliographystyle{spphys}       % APS-like style for physics
%\bibliography{D:/PHYSICS/TEX/BIBTEX/qttg}   % name your BibTeX data base

\begin{thebibliography}{100}
\providecommand{\url}[1]{{#1}}
\providecommand{\urlprefix}{URL }
\expandafter\ifx\csname urlstyle\endcsname\relax
  \providecommand{\doi}[1]{DOI \discretionary{}{}{}#1}\else
  \providecommand{\doi}{DOI \discretionary{}{}{}\begingroup
  \urlstyle{rm}\Url}\fi

\bibitem{Vining2009}
C.B. Vining, Nature Materials \textbf{8}, 83 (2009)

\bibitem{Snyder2008}
G.J. Snyder, E.S. Toberer, Nature Materials \textbf{7}, 105 (2008)

\bibitem{Minnich2009}
A.J. Minnich, M.S. Dresselhaus, Z.F. Ren, G.~Chen, Energy Environ. Sci.
  \textbf{2}, 466 (2009)

\bibitem{Mahan1996}
G.D. Mahan, J.O. Sofo, Proc. Natl. Acad. Sci. U.S.A. \textbf{93}, 7436 (1996)

\bibitem{Heremans2008}
J.P. Heremans, V.~Jovovic, E.S. Toberer, A.~Saramat, K.~Kurosaki,
  A.~Charoenphakdee, S.~Yamanaka, G.J. Snyder, Science \textbf{321}, 54 (2008)

\bibitem{Kim2009}
R.~Kim, S.~Datta, M.S. Lundstrom, Journal of Applied Physics \textbf{105}(3),
  034506 (2009).


\bibitem{Hochbaum2008}
A.I. Hochbaum, R.~Chen, R.D. Delgado, W.~Liang, E.C. Garnett, M.~Najarian,
  A.~Majumdar, P.~Yang, Nature \textbf{451}, 163 (2008)

\bibitem{Boukai2008}
A.I. Boukai, Y.~Bunimovich, J.~Tahir-Kheli, J.K. Yu, W.A.G. III, J.R. Heath,
  Nature \textbf{451}, 168 (2008)

\bibitem{Reddy2007}
P.~Reddy, S.Y. Jang, R.A. Segalman, A.~Majumdar, Science \textbf{315}, 1568
  (2007)

\bibitem{Baheti2008}
K.~Baheti, J.A. Malen, P.~Doak, P.~Reddy, S.Y. Jang, T.D. Tilley, A.~Majumdar,
  R.A. Segalman, Nano Letters \textbf{8}, 715 (2008).

\bibitem{Malen2009}
J.A. Malen, P.~Doak, K.~Baheti, T.D. Tilley, R.A. Segalman, A.~Majumdar, Nano
  Lett. \textbf{9}, 1164 (2009)

\bibitem{Malen2010}
J.A. Malen, S.K. Yee, A.~Majumdar, R.A. Segalman, Chem. Phys. Lett.
  \textbf{491}, 109 (2010)

\bibitem{Tan2010}
A.~Tan, S.~Sadat, P.~Reddy, Appl. Phys. Lett. \textbf{96}(1), 013110 (2010).

\bibitem{Hoffmann2009}
E.A. Hoffmann, H.A. Nilsson, J.E. Matthews, N.~Nakpathomkun, A.I. Persson,
  L.~Samuelson, H.~Linke, Nano Lett. \textbf{9}, 779 (2009)

\bibitem{Chowdhury2009}
I.~Chowdhury, R.~Prasher, K.~Lofgreen, G.~Chrysler, S.~Narasimhan, R.~Mahajan,
  D.~Koester, R.~Alley, R.~Venkatasubramanian, Nature Nanotechnology
  \textbf{4}, 235 (2009)

\bibitem{Dubi2011}
Y.~Dubi, M.~Di~Ventra, Rev. Mod. Phys. \textbf{83}(1), 131 (2011).

\bibitem{Bergfield2009}
J.P. Bergfield, C.A. Stafford, Nano Lett. \textbf{9}, 3072 (2009)

\bibitem{Kubala2008}
B.~Kubala, J.~K\"onig, J.~Pekola, Phys. Rev. Lett. \textbf{100}, 066801 (2008).

\bibitem{Held2009}
K.~Held, R.~Arita, V.~Anisimov, K.~Kuroki, arXiv:0903.2994  (2009)

\bibitem{Held2007}
K.~Held, Advances in Physics \textbf{56}, 829 (2007)

\bibitem{Wissgott2010}
P.~Wissgott, A.~Toschi, H.~Usui, K.~Kuroki, K.~Held, Phys. Rev. B \textbf{82},
  201106 (2010).


\bibitem{Boese2001}
D.~Boese, R.~Fazio, EPL (Europhysics Letters) \textbf{56}, 576 (2001)

\bibitem{Zhang2011}
Y.~Zhang, M.S. Dresselhaus, Y.~Shi, Z.~Ren, G.~Chen, Nano Lett. \textbf{11},
  1166 (2011)

\bibitem{Pauly2008}
F.~Pauly, J.K. Viljas, J.C. Cuevas, Phys. Rev. B \textbf{78}(3), 035315 (2008).

\bibitem{Ke2009}
S.H. Ke, W.~Yang, S.~Curtarolo, H.U. Baranger, Nano Letters \textbf{9}(3), 1011
  (2009).


\bibitem{Finch2009}
C.M. Finch, V.M. Garc\'\i{}a-Su\'arez, C.J. Lambert, Phys. Rev. B
  \textbf{79}(3), 033405 (2009).


\bibitem{Liu2009}
Y.S. Liu, Y.C. Chen, Phys. Rev. B \textbf{79}, 193101 (2009).

\bibitem{Liu2009a}
Y.S. Liu, Y.R. Chen, Y.C. Chen, ACS Nano \textbf{3}, 3497 (2009)

\bibitem{Liu2011a}
Y.S. Liu, H.T. Yao, Y.C. Chen, J. Phys. Chem. C \textbf{115}, 14988 (2011)

\bibitem{Quek2011}
S.Y. Quek, H.J. Choi, S.G. Louie, J.B. Neaton, ACS \textbf{5}, 551 (2011)

\bibitem{Nozaki2010}
D.~Nozaki, H.~Sevin\ifmmode~\mbox{\c{c}}\else \c{c}\fi{}li, W.~Li,
  R.~Guti\'errez, G.~Cuniberti, Phys. Rev. B \textbf{81}(23), 235406 (2010).

\bibitem{Saha2011}
K.K. Saha, T.~Markussen, K.S. Thygesen, B.K. Nikoli\ifmmode~\acute{c}\else
  \'{c}\fi{}, Phys. Rev. B \textbf{84}(4), 041412(R) (2011).


\bibitem{Sergueev2011}
N.~Sergueev, S.~Shin, M.~Kaviany, B.~Dunietz, Phys. Rev. B \textbf{83}, 195415
  (2011).


\bibitem{Bergfield2011}
J.P. Bergfield, G.C. Solomon, C.A. Stafford, M.A. Ratner, Nano Lett.
  \textbf{11}, 2759 (2011)

\bibitem{Murphy2008}
P.~Murphy, S.~Mukerjee, J.~Moore, Phys. Rev. B \textbf{78}(16), 161406 (2008).


\bibitem{Leijnse2010}
M.~Leijnse, M.R. Wegewijs, K.~Flensberg, Phys. Rev. B \textbf{82}(4), 045412
  (2010).


\bibitem{Entin-Wohlman2010}
O.~Entin-Wohlman, Y.~Imry, A.~Aharony, Phys. Rev. B \textbf{82}(11), 115314
  (2010).

\bibitem{Stadler2011}
R.~Stadler, T.~Markussen, J. Chem. Phys. \textbf{135}, 154109 (2011)

\bibitem{Markussen2009}
T.~Markussen, A.P. Jauho, M.~Brandbyge, Phys. Rev. Lett. \textbf{103}(5),
  055502 (2009).


\bibitem{Tsutsui2010a}
M.~Tsutsui, M.~Taniguchi, K.~Yokota, T.~Kawai, Appl. Phys. Lett.
  \textbf{96}(10), 103110 (2010).


\bibitem{Song2011}
H.~Song, M.A. Reed, T.~Lee, Adv. Mater. \textbf{23}, 1583 (2011)

\bibitem{Paulsson2003}
M.~Paulsson, S.~Datta, Phys. Rev. B \textbf{67}(24), 241403 (2003).


\bibitem{Cardamone2006}
D.~Cardamone, C.~Stafford, S.~Mazumdar, Nano Letters \textbf{6}(11), 2422
  (2006)

\bibitem{Ke2008}
S.H. Ke, W.~Yang, H.U. Baranger, Nano Letters \textbf{8}(10), 3257 (2008).

\bibitem{Saha2010}
K.K. Saha, B.K. Nikoli\'{c}, V.~Meunier, W.~Lu, J.~Bernholc, Phys. Rev. Lett.
  \textbf{105}(23), 236803 (2010).


\bibitem{Markussen2010a}
T.~Markussen, R.~Stadler, K.S. Thygesen, Nano Lett. \textbf{10}, 4260 (2010)

\bibitem{Markussen2011}
T.~Markussen, R.~Stadler, K.S. Thygesen, Phys. Chem. Chem. Phys. \textbf{13},
  14311 (2011)

\bibitem{Galperin2008}
M.~Galperin, A.~Nitzan, M.A. Ratner, Molecular Physics \textbf{106}, 397 (2008)

\bibitem{Frederiksen2007}
T.~Frederiksen, M.~Paulsson, M.~Brandbyge, A.P. Jauho, Phys. Rev. B
  \textbf{75}, 205413 (2007).


\bibitem{Dash2011}
L.K. Dash, H.~Ness, R.W. Godby, Phys. Rev. B \textbf{84}, 085433 (2011).

\bibitem{Mingo2006}
N.~Mingo, Phys. Rev. B \textbf{74}, 125402 (2006).


\bibitem{Vo2008}
T.T. Vo, A.J. Williamson, V.~Lordi, G.~Galli, Nano Lett. \textbf{8}, 1111
  (2008)

\bibitem{Jeong2010}
C.~Jeong, R.~Kim, M.~Luisier, S.~Datta, M.~Lundstrom, Journal of Applied
  Physics \textbf{107}(2), 023707 (2010).


\bibitem{Breuer2002}
H.P. Breuer, F.~Petruccione, \emph{The Theory of Open Quantum Systems} (Oxford
  University Press, Oxford, 2002)

\bibitem{Mitra2004}
A.~Mitra, I.~Aleiner, A.J. Millis, Phys. Rev. B \textbf{69}, 245302 (2004).

\bibitem{Dubi2008}
Y.~Dubi, M.~Di~Ventra, Nano Lett. \textbf{9}, 97 (2008)

\bibitem{Timm2008}
C.~Timm, Phys. Rev. B \textbf{77}, 195416 (2008).


\bibitem{Haug2007}
H.~Haug, A.P. Jauho, \emph{Quantum kinetics in transport and optics of
  semiconductors} (Springer, Berlin, 2007)

\bibitem{Haupt2010}
F.~Haupt, T.~Novotn\'{y}, W.~Belzig, Phys. Rev. B \textbf{82}, 165441 (2010).


\bibitem{Hartle2009}
R.~H\"artle, C.~Benesch, M.~Thoss, Phys. Rev. Lett. \textbf{102}, 146801
  (2009).

\bibitem{Datta1995}
S.~Datta, \emph{Electronic transport in mesoscopic systems} (Cambridge
  University Press, Cambridge, 1995)

\bibitem{Reich2002}
S.~{Reich}, J.~{Maultzsch}, C.~{Thomsen}, P.~{Ordej{\'o}n}, Physical Review B
  \textbf{66}(3), 035412 (2002).


\bibitem{Cresti2008}
A.~Cresti, N.~Nemec, B.~Biel, G.~Niebler, F.~Triozon, G.~Cuniberti, S.~Roche,
  Nano Research \textbf{1}, 361 (2008)

\bibitem{Markussen2009a}
T.~Markussen, A.P. Jauho, M.~Brandbyge, Phys. Rev. B \textbf{79}, 035415
  (2009).


\bibitem{Cervantes-Sodi2008}
F.~Cervantes-Sodi, G.~Cs\'anyi, S.~Piscanec, A.C. Ferrari, Phys. Rev. B
  \textbf{77}(16), 165427 (2008).


\bibitem{Areshkin2010}
D.A. Areshkin, B.K. Nikoli\'{c}, Phys. Rev. B \textbf{81}(15), 155450 (2010).


\bibitem{Cuniberti2005}
G.~Cuniberti, G.~Fagas, K.~Richter (eds.), \emph{Introducing molecular
  electronics} (Springer, Berlin, 2005)

\bibitem{Fiolhais2003}
C.~Fiolhais, F.~Nogueira, M.A. Marques (eds.), \emph{A Primer in Density
  Functional Theory (Lecture Notes in Physics Vol. 620)} (Springer Berlin,
  2003)

\bibitem{Taylor2001}
J.~{Taylor}, H.~{Guo}, J.~{Wang}, Physical Review B \textbf{63}(24), 245407
  (2001).


\bibitem{Brandbyge2002}
M.~{Brandbyge}, J.L. {Mozos}, P.~{Ordej{\'o}n}, J.~{Taylor}, K.~{Stokbro},
  Physical Review B \textbf{65}(16), 165401 (2002).


\bibitem{Stokbro2008}
K.~Stokbro, Journal of Physics: Condensed Matter \textbf{20}(6), 064216 (7pp)
  (2008).


\bibitem{Rungger2008}
I.~Rungger, S.~Sanvito, Phys. Rev. B \textbf{78}, 035407 (2008).


\bibitem{Saha2009a}
K.K. Saha, W.~Lu, J.~Bernholc, V.~Meunier, Journal of Chemical Physics
  \textbf{131}, 164105 (2009)

\bibitem{Esfarjani2006}
K.~Esfarjani, M.~Zebarjadi, Y.~Kawazoe, Phys. Rev. B \textbf{73}(8), 085406
  (2006).


\bibitem{Strange2011}
M.~Strange, C.~Rostgaard, H.~H\"akkinen, K.S. Thygesen, Phys. Rev. B
  \textbf{83}, 115108 (2011).


\bibitem{Wang2008c}
J.S. Wang, J.~Wang, J.T. L\"u, Eur. Phys. J. B \textbf{62}, 381 (2008)

\bibitem{McGaughey2004}
A.J.H. McGaughey, M.~Kaviany, Phys. Rev. B \textbf{69}, 094303 (2004).


\bibitem{McGaughey2006}
A.J.H. McGaughey, M.~Kaviany, Advances in Heat Transfer \textbf{39}, 169
  (Academic Press, San Diego, 2006)

\bibitem{Wang2006h}
R.Y. Wang, R.A. Segalman, A.~Majumdar, Appl. Phys. Lett. \textbf{89}(17),
  173113 (2006).


\bibitem{Rego1998}
L.G.C. Rego, G.~Kirczenow, Phys. Rev. Lett. \textbf{81}, 232 (1998).


\bibitem{gpaw}
https://wiki.fysik.dtu.dk/gpaw/

\bibitem{Enkovaara2010}
J.~Enkovaara, C.~Rostgaard, J.J. Mortensen, J.~Chen, M.~Dulak, L.~Ferrighi,
  J.~Gavnholt, C.~Glinsvad, V.~Haikola, H.A. Hansen, H.H. Kristoffersen,
  M.~Kuisma, A.H. Larsen, L.~Lehtovaara, M.~Ljungberg, O.~Lopez-Acevedo, P.G.
  Moses, J.~Ojanen, T.~Olsen, V.~Petzold, N.A. Romero, J.~Stausholm-Moller,
  M.~Strange, G.A. Tritsaris, M.~Vanin, M.~Walter, B.~Hammer, H.~Hakkinen,
  G.K.H. Madsen, R.M. Nieminen, J.K. Norskov, M.~Puska, T.T. Rantala,
  J.~Schiotz, K.S. Thygesen, K.W. Jacobsen, Journal of Physics: Condensed
  Matter \textbf{22}(25), 253202 (2010).


\bibitem{Ghosh2009}
S.~Ghosh, D.L. Nika, E.P. Pokatilov, A.A. Balandin, New J. Phys. \textbf{11},
  095012 (2009)

\bibitem{Seol2010}
J.H. Seol, I.~Jo, A.L. Moore, L.~Lindsay, Z.H. Aitken, M.T. Pettes, X.~Li,
  Z.~Yao, R.~Huang, D.~Broido, N.~Mingo, R.S. Ruoff, L.~Shi, Science
  \textbf{328}, 213 (2010)

\bibitem{Chen2011}
S.S. Chen, A.L. Moore, W.W. Cai, J.W. Suk, J.~An, C.~Mishra, C.~Amos,
  C.~Magnuson, J.~Kang, L.~Shi, R.S. Ruoff, ACS Nano \textbf{5}, 321 (2011)

\bibitem{Aksamija2011}
Z.~Aksamija, I.~Knezevic, Appl. Phys. Lett. \textbf{98}(14), 141919 (2011).


\bibitem{Cai2010}
J.~Cai, P.~Ruffieux, R.~Jaafar, M.~Bieri, T.~Braun, S.~Blankenburg, M.~Muoth,
  A.P. Seitsonen, M.~Saleh, X.~Feng, K.~Mullen, R.~Fasel, Nature
  \textbf{466}(7305), 470 (2010).


\bibitem{Jia2009}
X.~Jia, M.~Hofmann, V.~Meunier, B.G. Sumpter, J.~Campos-Delgado, J.~Manuel,
  R.H. Hyungbin, S.~Ya-Ping, H.A. Reina, J.~Kong, M.~Terrones, M.S.
  Dresselhaus, Science \textbf{323}, 1701 (2009)

\bibitem{Tao2011}
C.~Tao, L.~Jiao, O.V. Yazyev, Y.C. Chen, J.~Feng, X.~Zhang, R.B. Capaz,
  J.M.T.A. Zettl, S.G. Louie, H.~Dai, M.F. Crommie, Nature Phys. \textbf{7},
  616 (2011)

\bibitem{Ke2007}
S.H. Ke, H.U. Baranger, W.~Yang, Physical Review Letters \textbf{99}(14),
  146802 (2007).


\bibitem{Yazyev2008}
O.V. Yazyev, M.I. Katsnelson, Physical Review Letters \textbf{100}(4), 047209
  (2008).


\bibitem{Kunstmann2011}
J.~Kunstmann, C.~\"Ozdo\ifmmode~\breve{g}\else \u{g}\fi{}an, A.~Quandt,
  H.~Fehske, Phys. Rev. B \textbf{83}(4), 045414 (2011).


\bibitem{Prins2011}
F.~Prins, A.~Barreiro, J.W. Ruitenberg, J.S. Seldenthuis, N.~Aliaga-Alcalde,
  L.M.K. Vandersypen, H.S.J. van~der Zant, Nano Lett., DOI: 10.1021/nl202065x
  (2011)

\bibitem{Guo2006}
X.~{Guo}, J.P. {Small}, J.E. {Klare}, Y.~{Wang}, M.S. {Purewal}, I.W. {Tam},
  B.H. {Hong}, R.~{Caldwell}, L.~{Huang}, S.~{O'Brien}, J.~{Yan}, R.~{Breslow},
  S.J. {Wind}, J.~{Hone}, P.~{Kim}, C.~{Nuckolls}, Science \textbf{311}, 356
  (2006).

\bibitem{Zuev2009}
Y.M. Zuev, W.~Chang, P.~Kim, Phys. Rev. Lett. \textbf{102}(9), 096807 (2009).


\bibitem{Velev2004}
J.~Velev, W.~Butler, Journal of Physics: Condensed Matter \textbf{16}(21), R637
  (2004).


\bibitem{Lopez-Sancho1984}
M.P. Lopez-Sancho, J.M. Lopez-Sancho, J.~Rubio, J. Phys. F \textbf{14}, 1205
  (1984)

\bibitem{gulp}
http://projects.ivec.org/gulp/

\bibitem{Zimmermann2008}
J.~Zimmermann, P.~Pavone, G.~Cuniberti, Phys. Rev. B \textbf{78}, 045410
  (2008).

\bibitem{Sevincli2010}
H.~Sevin\ifmmode~\mbox{\c{c}}\else \c{c}\fi{}li, G.~Cuniberti, Phys. Rev. B
  \textbf{81}, 113401 (2010).


\bibitem{Brenner1990}
D.W. Brenner, Phys. Rev. B \textbf{42}, 9458 (1990).

\bibitem{Lindsay2010}
L.~Lindsay, D.A. Broido, Phys. Rev. B \textbf{81}, 205441 (2010).


\bibitem{Gale1997}
J.D. Gale, J. Chem. Soc., Faraday Trans. \textbf{93}, 629 (1997)

\bibitem{Galperin2007}
M.~Galperin, M.A. Ratner, A.~Nitzan, J. Phys.: Condens. Matter \textbf{19},
  103201 (2007)

\bibitem{Horsfield2006}
A.P. Horsfield, D.R. Bowler, H.~Ness, C.G. S\'{a}nchez, T.N. Todorov, A.J.
  Fisher, Rep. Prog. Phys. \textbf{69}, 1195 (2006)

\bibitem{Lu2007a}
J.T. L\"u, J.S. Wang, Phys. Rev. B \textbf{76}, 165418 (2007).

\bibitem{Hsu2011}
B.C. Hsu, Y.S. Liu, S.H. Lin, Y.C. Chen, Phys. Rev. B \textbf{83}, 041404
  (2011).


\bibitem{Asai2008}
Y.~Asai, Phys. Rev. B \textbf{78}, 045434 (2008).

\bibitem{Jiang2011}
J.W. Jiang, J.S. Wang, arXiv:1108.5817  (2011)

\bibitem{Choi2010}
W.S. Choi, H.~Ohta, S.J. Moon, Y.S. Lee, T.W. Noh, Phys. Rev. B \textbf{82},
  024301 (2010).

\bibitem{Scarola2002}
V.W. Scarola, G.D. Mahan, Phys. Rev. B \textbf{66}, 205405 (2002).


\end{thebibliography}

% Non-BibTeX users please use
%\begin{thebibliography}{}
%
% and use \bibitem to create references. Consult the Instructions
% for authors for reference list style.
%
%\bibitem{RefJ}
% Format for Journal Reference
%Author, Article title, Journal, Volume, page numbers (year)
% Format for books
%\bibitem{RefB}
%Author, Book title, page numbers. Publisher, place (year)
% etc
%\end{thebibliography}

\end{document}